\magnification=1200
\baselineskip=20 pt

\def\lijk{\lambda_{ijk}}
\def\lpijk{\lambda^{\prime}_{ijk}}
\def\lptojk{\lambda^{\prime}_{2jk}}
\def\lpojk{\lambda^{\prime}_{1jk}}
\def\lptrjk{\lambda^{\prime}_{3jk}}

\centerline{\bf Violation of supersymmetric equivalence in R parity}
\centerline{\bf violating couplings}

\vskip 1true in

\centerline{\bf Uma Mahanta}
\centerline{\bf Mehta Research Institute}
\centerline{\bf Chhatnag Road, Jhusi}
\centerline{\bf Allahabad-211019}
\centerline{\bf India}
\vskip .4true in
\centerline{\bf Abstract}

In this paper we consider the violation of supersymmetric equivalence
among the R parity violating couplings $\lijk $
 caused by  widely split
chiral supermultiplets. The calculations have been done for two specific
models of supersymmetry breaking: a) heavy SQCD models and b) 2-1 models.
We find that if $\lptojk \approx g$ and  $\lpojk \approx e$ then the 
violation of SUSY equivalence is of the order of 5-6 \% in heavy SQCD
models. On the other hand if  $\lptrjk \approx g$ and $\lijk \approx e$
then the violation of  of SUSY equivalence is of the order of 9.4 \%
in 2-1 models.

\vfill\eject

\centerline{\bf Introduction}

Low energy supersymmetry (SUSY) provides an attractive solution to the
naturalness problem  and perhaps also the heirarchy problem of the SM [1].
The search for SUSY will therefore constitute one of the major activities
of all future high energy colliders. After the discovery of SUSY the 
next task will be to measure the masses and couplings of the sparticles
with high precision. If SUSY is exact then the couplings that are related
by means of a supersymmetric transformation must be equal to all orders
in perturbation theory. However if SUSY is softly broken then although
these couplings will be equal at the tree level, radiative corrections
will introduce small splitting between them. Such violation of 
supersymmetric equivalence among the couplings increases logarithmically
with the heavy sparticle mass scale (M) [2]. Therefore if some of 
the sparticles
are very heavy then although they decouple from most low energy processes,
they nevertheless produce a non-decoupling effect through 
radiative corrections. In fact it could happen that some of the sparticles
are very heavy and
 inaccessible at the future high energy colliders, but we can probe 
their masses and couplings indirectly by precision measurement of such
SUSY breaking effects.

\centerline{\bf Implications of exact SUSY on R parity breaking couplings}

Consider the R parity violating couplings that break lepton number but
conserves baryon number [3]. Such effects are given by the following
Lagrangians:

$$\eqalignno{L_1 &=\lijk [{\tilde\nu}^i_L {\bar e}^k_R e^j_L+{\tilde e}^j_L
{\bar e}^k_R \nu^i_L+({\tilde e}^{k}_R)^* ({\bar \nu}^i_L)^c e^j_L\cr
& -(i\leftrightarrow j)]+h.c. &(1)\cr}$$

$$\eqalignno{L_2 &=\lpijk[{\tilde \nu}^i_L {\bar d}^k_R d^j_L+{\tilde d}^j_L
{\bar d}^k_R \nu^i_L+({\tilde d}^k_R)^*({\bar \nu}^i_L)^c d^j_L\cr
&-{\tilde e}^i_L{\bar d}^k_R u^j_L-{\tilde u}^j_L{\bar d}^k_R e^i_L
-({\tilde d}^k_R)^*({\bar e}^i_L)^c u^j_L]+h.c. &(2)\cr}$$

In this paper we shall consider the violation of SUSY equivalence among 
the couplings of $L_1$ that are related by supersymmetric transformations.
The reason being $\lijk$ unlike $\lpijk$ are amenable to precision 
measurements since they are free from QCD related uncertainties. Secondly
in many models the squarks turn out to be very heavy and they lie in the
Tev range. For such models the couplings $\lpijk$ cannot be directly 
measured at colliders operating in the few hundred Gev range.
Note that the couplings associated with the three terms ${\tilde \nu}^i_L
{\bar d}^k_R d^j_L$, ${\tilde d}^j_L {\bar d}^k_R \nu^i_L$ and
$({\tilde d}^k_R)^* ({\bar \nu}^i_L)^c d^j_L$ are related by supersymmetric
transformations. If SUSY is exact then their couplings should be equal
and this common coupling has been denoted in the above by $\lijk$.
However if SUSY is broken at some high energy scale M then the 
three couplings will
differ from each other at a low energy scale $\mu\ll M$.
In broken SUSY we shall denote the three couplings by $\lijk ^{(1)}$, 
$\lijk ^{(2)}$
and $\lijk ^{(3)}$. They satisfy the boundary condition $\lijk ^1 (M)=
\lijk ^2 (M)=\lijk ^3(M)=\lijk $ at the mass scale M for heavy sparticles.
The splitting
between the couplings at $\mu$ arises because the heavy sparticles decouple 
at M. If the members of each supermultiplet have the same mass and the 
radiative 
corrections due to all of them are taken into account then the three 
couplings will evolve in the same manner and there will be no difference
between them at $\mu$. The splitting between the couplings at $\mu$ 
due to widely split supermultiplets can 
therefore be estimated by considering only the loop diagrams that involve
 one or more heavy sparticles. In this paper the splitting between the 
three couplings will be calculated for two specific models of SUSY
breaking: a) heavy SQCD models and b) 2-1 models.

\centerline{\bf Radiative corrections to $\lijk$ in heavy SQCD models}

In heavy SQCD models like gauge mediated SUSY breaking [4] the colored 
squarks and the gluinos are much heavier than the corresponding ordinary 
particles.
 Therefore the wavefunction
and vertex renormalization constants for $\lijk ^{(1)}$, $\lijk ^{(2)}$
and $\lijk ^{(3)}$ that involve at least one heavy sparicle can arise from
$\lpijk$ only. Consider the vertex renormalization constants associated
with the three couplings. Vertex renormalization diagrams for $\lijk ^{(1)}$
that contain
at least one heavy squark line can arise  if $L_2$ contains a ${\tilde \nu}
q{\bar q}$ vertex, a ${\tilde q}^*{\bar e}_R q$ vertex and a ${\tilde q}
{\bar q} e_L$ vertx. However we find that although $L_2$ contains 
a ${\tilde \nu}^i
{\bar d}^k_R d^j_L$ vertex and a ${\tilde u}^j_L{\bar d}^k_R e^i_L$ vertex
it does not contain a ${\tilde q}{\bar e_R}q^{\prime}_L$ vertex. Hence there is
no relevant vertex correction diagram for $\lijk ^1$. Similarly there are 
also no vertex correction diagrams for $\lijk ^{(2)}$ and $\lijk ^{(3)}$ since
$L_2$ does not contain ${\tilde q}{\bar e}_R q^{\prime}_L$ and
${\tilde e}^*_R {\bar q} q^{\prime}$ vertices. In order that the ratios 
${\lijk ^{(1)}\over \lijk ^{(2)}}$ and
 ${\lijk^{(1)}\over \lijk ^{(3)}}$ measure the effects 
of SUSY breaking only we must ensure that the same set of family indices are 
involved in $\lambda^{(1)}$, $\lambda^{(2)}$ and $\lambda^{(3)}$. 
For example let us
consider the case of $i=2$ and $j=k=1$. In evaluating the renormalization
constants we shall use the $\overline {MS}$ scheme and 
keep only the leading log terms and the terms that remain
finite in the limit ${\mu \over M}\rightarrow 0$ where $\mu$ is the 
mass scale for
light sparticles. Consider now the wavefunction renormalization constants 
associated with $\lambda^{(1)}_{211}$, $\lambda^{(2)}_{211}$ and 
$\lambda^{(3)}_{211}$.
We find that only the self energy diagrams for $e^1_L$, $\nu^2_L$ and
$\nu^{2c}_L$ involves a heavy squark line. The wavefunction renormalization
constants are given by

$${\sqrt Z}_{e^1_L}=1-{N_c\over 64\pi^2}\sum_{j,k}
\vert \lpojk\vert^2 (\ln{M^2\over \mu^2}-1).\eqno(3)$$.

and

$${\sqrt Z}_{\nu^2_L}={\sqrt Z}_{\nu^{2c}_L}=
1-{N_c\over 64\pi^2}\sum_{jk}\vert \lambda^{\prime}_{2jk}\vert^2 
(\ln{M^2\over \mu^2}-1).\eqno(4)$$.

where $N_c=3$ is the number of colors and $\mu$ is the renormalization
mass scale which we shall assume to be around 100 Gev. Using the
boundary condition $\lambda^1_{211}(M)=\lambda^2_{211}(M)=\lambda^3_{211}(M)$
we then get

$${\lambda^{(2)}_{211}(\mu )\over \lambda^{(1)}_{211}(\mu )}=
1-{N_cN^2_g\over 64 
\pi^2}({\bar \lambda}^{\prime 2}_2-{\bar \lambda}^{\prime 2}_1)(\ln {M^2
\over \mu^2}-1).\eqno(5)$$
and 
$${\lambda^{(3)}_{211}(\mu )\over \lambda^{(1)}_{211}(\mu )}=
1-{N_cN^2_g\over 64 
\pi^2}{\bar \lambda}^{\prime 2}_2(\ln {M^2
\over \mu^2}-1).\eqno(6)$$

In the above $N_g^2{\bar \lambda}^{\prime 2}_1 \equiv \sum_{j,k}
\vert \lambda^{\prime}_{1jk}\vert^2 $ and 
$N_g^2{\bar \lambda}^{\prime 2}_2 \equiv \sum_{j,k}
\vert \lambda^{\prime}_{2jk}\vert^2 $. $N_g$ is the number of fermion
generations. In models of gauge mediated SUSY breaking if the light 
sparticles have a mass of few hundred Gev then the squarks usually 
lie in the 1 Tev mass range. For ${\bar \lambda}^{\prime 2}_1 =e^2$
and ${\bar \lambda}^{\prime 2}_2 =g^2$ we find that

$$\lambda^{(1)}_{211}(\mu ):\lambda^{(2)}_{211}(\mu ) :
\lambda^{(3)}_{211}(\mu )=1:1-.051:1-.066 \eqno(7)$$

So in this case the violation of SUSY equivalence is 5.1\% between 
$\lambda^{(1)}_{211}$ and $\lambda^{(2)}_{211}$ and 6.6\% between
$\lambda^{(3)}_{211}$ and $\lambda^{(1)}_{211}$. The splittings 
between $\lambda^{(1)}_{211}$,  $\lambda^{(2)}_{211}$ and
$\lambda^{(3)}_{211}$ in heavy SQCD models
is therefore quite large to be detectable
through precision measurements of $\lijk$ at future $e^+e^-$
colliders.

The constraints on $\lpijk$ derived from low energy phenomenology
that are given in standard references [5]
scale with the sparticle mass scale. The couplings $\lpijk$ that appear in
the above radiative corrections involve one heavy squark field in the Tev
range.
The present
experimental bounds on such couplings are therefore very weak. The reason is 
not hard to see. Sparticles with a mass of around 1 Tev decouple from low 
energy processes and therefore their R parity violating couplings can 
be quite large. The values of
${\bar \lambda}^{\prime}_1$ and ${\bar \lambda}^{\prime}_2$ assumed above 
are therefore
consistent with the experimental bounds and in fact are much more 
restrictive. If we had used the present experimental bounds on $\lpijk$
the splitting between  $\lambda^{(1)}_{211}$,  $\lambda^{(2)}_{211}$ and
$\lambda^{(3)}_{211}$ would have been much larger.

\centerline{\bf Radiative corrtections to $\lijk$ in 2-1 models}

In 2-1 models [6] all sparticles belonging to the first and second generations
 are very heavy. So in this case we also have to include the 
radiative corrections from loop diagrams that contain one heavy slepton line
belonging to the first or second generation. Since only the sleptons of 
the third generation are light consider the following terms of $L_1$:
$$\eqalignno{L_1&=\lambda^{(1)}_{3jk}{\tilde\nu}^3_L{\bar e}^k_Re^j_L+
\lambda^{(2)}_{i3k}{\tilde e}^3_L{\bar e}^k_R\nu^i_L\cr
&+\lambda^{(3)}_{ij3}{\tilde e}^{3*}_R{\bar \nu}^{ic}_Le^j_L+..&(8)\cr}$$

Note that since $\lijk$ must be antisymmetric in the first two indices
it is not possible to compare $\lambda^{(1)}_{3jk}$ with 
$\lambda^{(2)}_{i3k}$ as a measure of SUSY breaking. The only possibilities 
are to compare $\lambda^1_{3j3}$ with $\lambda^{(3)}_{3j3}$ or 
$\lambda^{(2)}_{i33}$ with $\lambda^{(3)}_{i33}$. Let us consider
for example  $\lambda^1_{323}$ and $\lambda^{(3)}_{323}$. As before there 
is no vertex correction associated with  $\lambda^1_{323}$ or 
$\lambda^{(3)}_{323}$ arising from $\lpijk$. It can also be shown that
there is no vertex correction for  $\lambda^1_{323}$ or $\lambda^{(3)}_{323}$
arising from $\lijk$ that involves a heavy slepton line.
The renormalizations of  $\lambda^1_{323}$ and $\lambda^{(3)}_{323}$ 
from M to $\mu$ are therefore given solely by their respective 
wavefunction renormalization constants. We find that in 2-1 models the
wavefunction renormalization of $e^3_R$ and $\nu^{3c}_L$ are given by

$${\sqrt Z}_{e^3_R}=[1-{1\over 32\pi^2}(2\vert \lambda_{123}\vert^2
+\vert\lambda_{233}\vert^2+\vert \lambda_{133}\vert ^2)
(\ln{M^2\over \mu^2}-1)].\eqno(9)$$
and 
$$\eqalignno{{\sqrt Z}_{\nu^{3c}_L}&=[1-{N_c\over 64\pi^2}\sum_{j=1}^3
\sum_{k=1}^2\vert \lambda^{\prime}_{3jk}\vert ^2(\ln{M^2\over \mu^2}-1)\cr
&-{1\over 64\pi^2}\sum_{j=1}^2\sum_{k=1}^2\vert \lambda_{3jk}\vert ^2
(\ln{M^2\over \mu ^2}-1)]&(10)\cr}$$

Note that in ${\sqrt Z}_{\nu^{3c}_L}$ the contribution from $\lpijk$
is summed over k from 1 to 2 since only the squarks of first two 
generations are heavy. Whereas in the contribution from $\lijk$ the sum 
over j and k are determined by the antisymmetry of $\lijk$ in the 
first two indices and the fact that only the sparticles of the third 
generation are light.

$$\eqalignno{{\lambda^{(1)}_{323}(\mu )\over \lambda^{(3)}_{323}(\mu)}&=1+
{1\over 64\pi^2}[N_c\sum_{j=1}^3\sum_{k=1}^2\vert \lambda^{\prime}_{3jk}
\vert^2+\sum_{j=1}^2\sum_{k=1}^2\vert\lambda_{3jk}\vert^2\cr
&-2(2\vert\lambda_{123}\vert^2+\vert\lambda_{233}\vert^2+
\vert\lambda_{133}\vert^2)](\ln{M^2\over \mu^2}-1)&(11)\cr}$$

In 2-1 models if the light sparticles are in the few
hundred Gev range then the heavy sparticles of the first two generations 
can lie in the 10 Tev range without violating the low energy constraints
arising from FCNC. To get a numerical estimate of the violation of 
SUSY equivalence between $\lambda^{(1)}_{323}$ and $\lambda^{(3)}_{323}$
let us assume that $\lambda^{\prime}_{3jk}\approx g$ and $\lambda_{1jk}
\approx e$. We then find that ${\lambda^{(1)}_{323}(\mu )\over
\lambda^{(3)}_{323}(\mu )}\approx 1+.094 $ which is again large enough 
to be detectable through precision measurements of $\lijk$ at future
$e^+e^-$ colliders.

\centerline{\bf Conclusions}

In this paper we have considered the violation of SUSY equivalence among 
the couplings $\lambda^{(1)}_{ijk}$, $\lambda^{(2)}_{ijk}$ and 
$\lambda^{(3)}_{ijk}$ that are related by means of supersymmetric 
transformations. We have computed the splitting between these couplings 
in the context of heavy SQCD models  and 2-1 models. We find that if
$\lambda^{\prime}_{2jk}\approx g$ and $\lambda^{\prime}_{1jk}\approx e$
then the violation of SUSY equivalence is of the order of 5-6\% in heavy 
SQCD models. On the other hand if $\lambda^{\prime}_{3jk}\approx g$
and $\lijk \approx e$ then the violation of SUSY equivalence can be as 
large as 9.4\% in 2-1 models. In either model the splittings
caused by widely split chiral supermultiplets 
 are quite large to be detectable through
precision measurements of $\lambda^{(1)}_{ijk}$, $\lambda^{(2)}_{ijk}$ 
and $\lambda^{(3)}_{ijk}$ at future $e^+e^-$ colliders.

\centerline{\bf References}

\item{1.} J. Ellis, K. Enqvist, D. Nanopoulos and F. Zwirner, Mod. Phys. Lett.
A 1, 57 (1986); R. Barbieri and G. F. Giudice, Nucl. Phys. B 306, 63 (1988);
G. W. Anderson and D. J. Castano, Phys. Lett. B 347, 300 (1995);
Phys. Rev. D 52, 1693 (1995).

\item{2.} H. C. Cheng, J. L. Feng and N. Polonsky, Phys. Rev. D 56, 6875
(1997); H. C. Cheng, J. L. Feng and N. Polonsky, ibid. 57, 152 (1998);
L. Randall, E. Katz and S. F. Su, hep-ph/9801416.

\item{3.} V. Barger, G. F. Giudice and T. Han, Phys. Rev. D 400, 2987 
(1989).

\item{4.} M. Dine and  A. Nelson, Phys. Rev. D 48, 1277 (1993);
M. Dine, A. Nelson and Y. Shirman, ibid. 51, 1362 (1995).

\item{5.} H. Dreiner, hep-ph/9707435.

\item{6.} A. G. Cohen, D. B. Kaplan and A. E. Nelson, Phys. Lett. B 388, 
588 (1996); A. G. Cohen, D. B. Kaplan, F. Lepintre and A. E. Nelson, Phys.
Rev. Lett. 78, 2300 (1997).

\end